\documentclass[showkeys,superscriptaddress,floatfix,prd,8pt,aps]{revtex4-2}
\usepackage{graphicx,epstopdf}
\usepackage[caption=false]{subfig}
\usepackage{appendix}
\usepackage[T1]{fontenc}
\usepackage{lmodern}
\usepackage[dvipsnames,x11names]{xcolor}
\usepackage[colorlinks=true,linkcolor=ForestGreen,citecolor=ForestGreen,urlcolor=NavyBlue]{hyperref}
\usepackage[sort&compress]{natbib}
\usepackage{lipsum}
\usepackage{morefloats}
\usepackage[pdf]{pstricks}
\usepackage{amsmath}
\usepackage{amssymb}
\usepackage{amsfonts}
\usepackage{rotating}
\usepackage{cancel}
\usepackage{mathtools}
\usepackage{bbm}
\usepackage{dsfont}
\usepackage{bbold}
\usepackage{multirow}
\usepackage{ulem}
\usepackage{physics}
\usepackage{orcidlink}
\usepackage{colortbl}

\def\pslash{p\!\!\!\slash }
\def\qslash{q\!\!\!\slash }
\def\xslash{x\!\!\!\slash }
\def\eslash{\varepsilon\!\!\!\slash }

\def\vel{\left|}
\def\ver{\right|}

\begin{document}

\title{Magnetic dipole moments of the $\Omega_{c}(3185)^0$ and $\Omega_c(3327)^0$ states from molecular perspective
}

\author{Ula\c{s}~\"{O}zdem\orcidlink{0000-0002-1907-2894}}%
\email[]{ulasozdem@aydin.edu.tr }
\affiliation{ Health Services Vocational School of Higher Education, Istanbul Aydin University, Sefakoy-Kucukcekmece, 34295 Istanbul, T\"{u}rkiye}

\date{\today}
 
\begin{abstract}
In this paper we study the magnetic dipole moments of the newly discovered $\Omega_{c}(3185)^0$ and $\Omega_c(3327)^0$, assuming that $\Omega_{c}(3185)^0$ and $\Omega_c(3327)^0$ are $S$-wave $D \Xi$ and $D^* \Xi$ molecular pentaquark states, respectively.   Together with these states, the magnetic dipole moments of possible $D_s \Xi$, $D_s^* \Xi$, $D \Xi^*$, and $D_s \Xi^*$ pentaquark states are also studied.  The magnetic dipole moments of these singly-charmed pentaquarks are estimated within the framework of the QCD light cone sum rules utilizing the photon distribution amplitudes.  In the search for the properties of singly-charmed molecular pentaquark states, the results obtained for the magnetic dipole moments can be useful.
\end{abstract}

\maketitle

\section{Motivation}
 
In recent years there have been many experimental discoveries of excited baryon states. Deciphering and understanding the internal organization of excited baryon states remains a significant challenge in non-perturbative QCD. Discovering many new heavy baryons at experimental facilities requires identifying and classifying them  \cite{Guo:2017jvc,Olsen:2017bmm,Brambilla:2019esw,Chen:2022asf}. 
Theoretical studies contain the analysis of baryons containing a single heavy quark, which allows an ideal platform to explore the dynamics of a light diquark in the medium of a heavy quark, to test the predictions of various theoretical models, and to improve the understanding of the non-perturbative structure of QCD.
 In recent years, there have been considerable experimental breakthroughs in the sector of singly heavy baryons~ \cite{ARGUS:1993vtm, CLEO:1994oxm, ARGUS:1997snv, E687:1993bax, CLEO:2000mbh, BaBar:2006itc, Belle:2006xni, LHCb:2017jym, Ammosov:1993pi, CLEO:1996czm, Belle:2004zjl, CLEO:1995amh, CLEO:1996zcj,E687:1998dwp, CLEO:2000ibb, CLEO:1999msf, LHCb:2020iby, BaBar:2007xtc, Belle:2006edu, BaBar:2007zjt, BaBar:2006pve, LHCb:2017uwr, LHCb:2012kxf, LHCb:2020lzx, LHCb:2019soc, LHCb:2018haf, CMS:2021rvl,  LHCb:2018vuc,  LHCb:2020xpu, LHCb:2020tqd}. Since the evidence for the existence of some of these states is weak and the quantum numbers are not well determined, further experimental research is needed. Thus, they continue to be studied in both experimental and theoretical research.
 
Two new excited singly-charmed states, $\Omega_c(3185)^0$ and $\Omega_c(3327)^0$, have recently been reported by the LHCb Collaboration in the $\Xi_c^+K^-$ invariant mass distribution \cite{LHCb:2023sxp}. The measured masses and decay widths are as follows 
\begin{align}
& M_{\Omega_c(3185)} = 3185.1\pm1.7^{+7.4}_{-0.9}\pm0.2 \mbox{ MeV}\,,~~~
\Gamma_{\Omega_c(3185)} = 50\pm7 ^{+10}_{-20}\mbox{ MeV} \,, \\
& M_{\Omega_c(3327)} = 3327.1\pm1.2^{+0.1}_{-1.3} \pm0.2\mbox{ MeV}\,,~~~
\Gamma_{\Omega_c(3327)} = 20\pm5 ^{+13}_{-1.0}\mbox{ MeV} \, .
\end{align}

 While the spin-parity quantum numbers of these states are waiting to be assigned, their observations could yield new information about the strong interaction and complex hadron spectroscopy for a better understanding of the underlying principles. 
The experimental discoveries have served as the basis for several theoretical studies, which, to shed light on their underlying structure, have investigated them in ordinary baryon~\cite{Pan:2023hwt,Luo:2023sra,Ortiz-Pacheco:2023kjn,Karliner:2023okv,Wang:2023wii,Yu:2023bxn} and molecular pentaquark states~\cite{Xin:2023gkf,Yan:2023ttx,Feng:2023ixl,Yan:2023tvl}.   
In Ref. \cite{Pan:2023hwt}, the masses of the $\Omega_{c}(3185)^0$ and $\Omega_c(3327)^0$ states in the quark-diquark configuration were obtained with the linear Regge trajectory model, and they found that the $\Omega_{c}(3185)^0$ state is a $2S$ state and the $\Omega_c(3327)^0$ state is a $1D$ state with spin-parity quantum numbers $J^P =1/2^+$ and $J^P =3/2^+$, respectively.  
In Ref. \cite{Luo:2023sra}, the authors have calculated the mass of the $\Omega_c(3327)^0$ state in the framework of the non-relativistic model using the Gaussian expansion method, and their results support that the $\Omega_c(3327)^0$ state is a good candidate for the $1D$ state with $J^P =5/2^+$.  
In Ref. \cite{Ortiz-Pacheco:2023kjn}, the mass of the $\Omega_{c}(3185)^0$ state was obtained employing the non-relativistic harmonic oscillator model, and the result of this study suggests that the $\Omega_{c}(3185)^0$ state is a good candidate for the $P$-wave $\rho$-excited state with $J^P =3/2^-$. 
In Ref. \cite{Karliner:2023okv}, the assignment of the $\Omega_{c}(3185)^0$ and $\Omega_c(3327)^0$ states was discussed. According to their results, the $1S-2S$ and hyperfine splittings are smaller and larger, respectively, than expected. However, the deviations from the predicted values are not large enough to jeopardize these assignments. 
In Ref. \cite{Wang:2023wii}, the mass of the $\Omega_c(3327)^0$ state was obtained in QCD sum rules, and this state is assigned to the $D$-wave $\Omega_c$ state with $J^P =1/2^+$, $J^P =3/2^+$ and $J^P =5/2^+$. 
In Ref. \cite{Yu:2023bxn}, the masses of the $\Omega_{c}(3185)^0$ and $\Omega_c(3327)^0$ states were obtained in the $^3P_0$ model, and the results of this study support the assignment of the $\Omega_{c}(3185)^0$ and $\Omega_c(3327)^0$ states as $2S(3/2^+)$ and $1D(3/2^+)$, respectively.    
%%%%%%%%%%%%%%%%%%%%%%%%%%%%%%%%%%%%%%%%%%%%%%%%%%%%%%%%%%%%%%%%%%%%%%%%%%%%%%%%%%%%%%%%%%%%%%%%%%%%%%%%%%%%
In Ref. \cite{Xin:2023gkf}, the QCD sum rules have been applied to study the mass of the $\Omega_{c}(3185)^0$ and $\Omega_c(3327)^0$ states, assuming that these states are molecular pentaquarks. Their results imply that 
the $\Omega_{c}(3185)^0$ is $D \Xi$ with $J^P =1/2^-$, and the $\Omega_c(3327)^0$ is $D^* \Xi$ with $J^P =3/2^-$. 
In Ref. \cite{Yan:2023ttx}, the authors used a phenomenological model to obtain the mass of the $\Omega_{c}(3185)^0$ and $\Omega_c(3327)^0$ states. They predicted that the $\Omega_{c}(3185)^0$ is the $D \Xi$ molecular state with $J^P =1/2^-$, and the $\Omega_c(3327)^0$ is the $D^* \Xi$ molecular state with $J^P =3/2^-$.  
In Ref. \cite{Feng:2023ixl}, the strong decays of the $\Omega_{c}(3185)^0$ and $\Omega_c(3327)^0$ states are predicted assuming the $\Omega_{c}(3185)^0$ and $\Omega_c(3327)^0$ as $S$-wave $D \Xi$ and $D^* \Xi$ molecular states within the effective Lagrangian method. Their predictions support the  $\Omega_c(3327)^0$ as a $J^P =3/2^-$ $D^* \Xi$ molecular state, and the $\Omega_{c}(3185)^0$ may be a meson-baryon molecule with a large $D \Xi$ component.   
In Ref. \cite{Yan:2023tvl}, the quark delocalization color screening model was used to obtain the mass of the $\Omega_{c}(3185)^0$ state. According to the results of this study, the $\Omega_{c}(3185)^0$ can be explained as the molecular $D \Xi$ with $J^P =1/2^-$.

%If one examines the studies in the literature listed above, it can be seen that almost all the calculations are aimed at calculating the spectroscopic parameters of these states, and it is easy to see that the spectroscopic parameters alone are not sufficient to elucidate the controversial nature of these states. Hence, it is obvious that further analyses such as the electromagnetic form factors, the weak decays, and so on are needed to shed light on the internal structure of these states. 
%
Although a great deal of effort has gone into their study, the $\Omega_{c}(3185)^0$ and $\Omega_c(3327)^0$ states are still not fully elucidated.  All of these studies with alternative structures, whose predictions are within the error of the experimental discoveries, imply that more work needs to be done on the $\Omega_{c}(3185)^0$ and $\Omega_c(3327)^0$ states. Thus, we need to continue to investigate these states, such as the weak decays, the electromagnetic form factors, etc., to better understand their characteristics. The electromagnetic form factors, especially the magnetic dipole moments, are of particular interest as they provide insight into the distribution of charge and magnetism within hadrons. 
The states $\Omega_{c}(3185)^0$ and $\Omega_c(3327)^0$ are close to the thresholds of $D \Xi$ and $D^* \Xi$, suggesting that they could be the molecular states of $D \Xi$ and $D^* \Xi$.   Because of this assumption, we consider the states $\Omega_{c}(3185)^0$ and $\Omega_c(3327)^0$  to be the bound states $D\Xi$ and $D^*\Xi$: $\Omega_{c}(3185)^0=(\mid D^0 \Xi^0 \rangle \, - \mid D^+ \Xi^- \rangle )/ \sqrt{2}$, and $\Omega_c(3327)^0=(\mid D^{*0} \Xi^0 \rangle \, - \mid D^{*+} \Xi^- \rangle )/ \sqrt{2}$; and we study the magnetic dipole moments (MDMs) of these states in the framework of QCD light-cone sum rules. 
Besides these states, the MDMs of possible $D_s \Xi$, $D_s^* \Xi$, $D \Xi^*$, and $D_s \Xi^*$  pentaquarks have also been analyzed. 
As known, QCD light-cone sum rules have been widely used to calculate masses, form factors, MDMs, decay constants,  etc. of conventional hadrons and are a robust tool for studying exotic hadron properties. The correlation function is calculated by the QCD light cone sum rules prescription concerning both hadrons (the hadronic side) and  quark-gluon degrees of freedom (the QCD side).
Then, the effects of the continuum and higher states are removed by applying the Borel transform and continuum subtraction procedure. As a final step, the physical quantities, i.e. the MDMs, are then obtained by matching these two different descriptions of the correlation function via quark-hadron duality~\cite{Chernyak:1990ag,Braun:1988qv,Balitsky:1989ry}.

The paper is organized as follows. In Sect. \ref{secII}, we construct the QCD light-cone sum rules for the MDMs of the molecular states with procedures that are similar to those used in our previous studies~\cite{Ozdem:2021ugy,Ozdem:2023htj,Ozdem:2022kei,Ozdem:2021vry,Ozdem:2021hmk,Ozdem:2022vip,Ozdem:2023eyz}. In Sect. \ref{secIII}, we illustrate our numerical results;  %This section also contains the discussions and a summary of the present study. %, 
and Sect. \ref{secIV} is dedicated to the summary and concluding remarks.
 The analytical formulas obtained for the MDM of the $\Omega_{c}(3185)^0$ state are provided in the Appendix.

 \begin{widetext}
  
\section{QCD light-cone sum rules for the magnetic dipole moments of the singly-charmed molecular pentaquark states}\label{secII}

 The first step in the analysis of MDMs of the spin-1/2 and spin-3/2 singly-charmed molecular pentaquark states (hereafter, $M_c$ and $M_c^*$, respectively) with the QCD light-cone sum rules is the introduction of the subsequent correlation functions,
 \begin{align} \label{edmn01}
\Pi(p,q)&=i\int d^4x e^{ip \cdot x} \langle0|T\left\{J^{M_c}(x)\bar{J}^{M_c}(0)\right\}|0\rangle _\gamma \, , \\
%\end{eqnarray}
%\begin{eqnarray} 
\Pi_{\mu\nu}(p,q)&=i\int d^4x e^{ip \cdot x} \langle0|T\left\{J_\mu^{M_c^*}(x)\bar{J}_\nu^{M_c^*}(0)\right\}|0\rangle _\gamma \,, \label{Pc101}
\end{align}
where $T$ is the time ordered product, sub-indice $\gamma$ is the external electromagnetic field. The $J^{M_c}(x)$ and $J^{P^*_{c}}(x)$ are the interpolating currents for the spin-$ 1/2^-$ and spin-$ 3/2^-$ states, respectively. These currents are necessary for further analysis and have been provided as 
\begin{align}\label{curpcs1}
 %%%%%%%%%%%%%%%%%%%%%%%%%%%%%%%%%%%%%%%%%%%%%%%%%%%%%%%%%
  %%%%%%%%%%%%%%%%%%%%%%%%%%%%%%%%%%%%%%%%%%%%%%%%%%%%%%%%%
J^{D_s \Xi}(x)& =\mid  D_s^+ \Xi^{0/-} \rangle  
=  \big[\bar s^d(x)i \gamma_5 c^d(x)\big]\big[\varepsilon^{abc} s^{a^T}(x)C\gamma_\mu s^b(x)   \gamma^\mu\gamma_5 q^c(x)\big],\\
%%%%%%%%%%%%%%%%%%%%%%%%%%%%%%%%%%%%%%%%%%%%%%%%%%%%%%%%%
%%%%%%%%%%%%%%%%%%%%%%%%%%%%%%%%%%%%%%%%%%%%%%%%%%%%%%%%%
%%%%%%%%%%%%%%%%%%%%%%%%%%%%%%%%%%%%%%%%%%%%%%%%%%%%%%%%%
J_\mu^{D_s^* \Xi}(x)&= \mid  D_s^{*+} \Xi^{0/-} \rangle =\big[\bar s^d(x) \gamma_\mu c^d(x)\big]\big[\varepsilon^{abc} s^{a^T}(x)C\gamma_\alpha    s^b(x) \gamma^\alpha  \gamma_5 q^c(x)\big]\,,\\
%%%%%%%%%%%%%%%%%%%%%%%%%%%%%%%%%%%%%%%%%%%%%%%%%%%%%%%%%
%%%%%%%%%%%%%%%%%%%%%%%%%%%%%%%%%%%%%%%%%%%%%%%%%%%%%%%%%
J_\mu^{D_s \Xi^{*}}(x)&= \mid  D_s^+ \Xi^{*0/-} \rangle =\big[\bar s^d(x) i \gamma_5 c^d(x)\big]\big[\varepsilon^{abc} s^{a^T}(x)C\gamma_\mu   s^b(x)   q^c(x)\big]\,,
 \end{align}
\begin{align}\label{curpcs2}
%%%%%%%%%%%%%%%%%%%%%%%%%%%%%%%%%%%%%%%%%%%%%%%%%%%%%%%%%
%%%%%%%%%%%%%%%%%%%%%%%%%%%%%%%%%%%%%%%%%%%%%%%%%%%%%%%%%
J^{D \Xi}(x)& =\frac{1}{\sqrt{2}}\Big \{\mid D^0 \Xi^0 \rangle \, - \mid D^+ \Xi^- \rangle  \Big \}=\frac{1}{\sqrt{2}} \Big \{ \big[\bar u^d(x)i \gamma_5 c^d(x)\big]\big[\varepsilon^{abc} s^{a^T}(x)C\gamma_\mu s^b(x)  \gamma^\mu\gamma_5 u^c(x)\big]\nonumber\\
&
 + \big[\bar d^d(x)i \gamma_5 c^d(x)\big]  
 \big[\varepsilon^{abc} s^{a^T}(x) C\gamma_\mu s^b(x) 
 \gamma^\mu\gamma_5 d^c(x)\big] \Big\} \, , \\
%%%%%%%%%%%%%%%%%%%%%%%%%%%%%%%%%%%%%%%%%%%%%%%%%%%%%%%%%
J_\mu^{D^* \Xi}(x)&=\frac{1}{\sqrt{2}}\Big \{\mid D^{*0} \Xi^0 \rangle \, - \mid D^{*+} \Xi^- \rangle  \Big \}=\frac{1}{\sqrt{2}} \Big \{ \big[\bar u^d(x) \gamma_\mu c^d(x)\big]\big[\varepsilon^{abc} s^{a^T}(x)C\gamma_\alpha s^b(x)  \gamma^\alpha\gamma_5 u^c(x)\big]
\nonumber\\
&  + 
\big[\bar d^d(x) \gamma_\mu c^d(x)\big]  
 \big[\varepsilon^{abc} s^{a^T}(x)
  C\gamma_\alpha s^b(x) 
 \gamma^\alpha\gamma_5 d^c(x)\big] \Big\} \,,\\  
%%%%%%%%%%%%%%%%%%%%%%%%%%%%%%%%%%%%%%%%%%%%%%%%%%%%%%%%%
J_\mu^{D \Xi^*}(x)&=
\frac{1}{\sqrt{2}}\Big \{\mid D^{0} \Xi^{*0} \rangle \, - \mid D^{+} \Xi^{*-} \rangle  \Big \}
=\frac{1}{\sqrt{2}} \Big \{ \big[\bar u^d(x) i \gamma_5 c^d(x)\big]\big[\varepsilon^{abc} s^{a^T}(x)C\gamma_\mu    s^b(x) u^c(x)\big] 
\nonumber\\
& + 
\big[\bar d^d(x) i \gamma_5 c^d(x)\big]  
 \big[\varepsilon^{abc} s^{a^T}(x) 
  C\gamma_\mu  s^b(x)   d^c(x)\big] \Big\} \,, 
  %%%%%%%%%%%%%%%%%%%%%%%%%%%%%%%%%%%%%%%%%%%%%%%%%%%%%%%%%
\end{align}
where  $q=u$ or $d$-quark, $a$, $b$, $c$ and  $d$ are color indices and the $C$ is the charge conjugation operator.

We initiate our analysis by obtaining the correlation functions for the hadronic representation. After inserting a complete set of intermediate states with the same content and quantum numbers of these molecular pentaquarks, and integrating over x, we derive the results
 \begin{align}\label{edmn02}
\Pi^{Had}(p,q)&=\frac{\langle0\mid J^{M_c}(x) \mid
{M_c}(p, s) \rangle}{[p^{2}-m_{M_c}^{2}]}
\langle {M_c}(p, s)\mid
{M_c}(p+q, s)\rangle_\gamma 
\frac{\langle {M_c}(p+q, s)\mid
\bar J^{M_c}(0) \mid 0\rangle}{[(p+q)^{2}-m_{M_c}^{2}]}+ \cdots , \\
%\end{align}
\nonumber\\ 
%\begin{align}
\Pi^{Had}_{\mu\nu}(p,q)&=\frac{\langle0\mid  J_{\mu}^{M_c^*}(x)\mid
{M_c^*}(p,s)\rangle}{[p^{2}-m_{{M_c^*}}^{2}]}
\langle {M_c^*}(p,s)\mid
{M_c^*}(p+q,s)\rangle_\gamma 
\frac{\langle {M_c^*}(p+q,s)\mid
\bar{J}_{\nu}^{M_c^*}(0)\mid 0\rangle}{[(p+q)^{2}-m_{{M_c^*}}^{2}]}+ \cdots .\label{Pc103}
\end{align}

The matrix elements in Eqs.~(\ref{edmn02}) and (\ref{Pc103}) can be formulated by expressing them in terms of hadronic parameters such as form factors, residues, etc., in the following manner,
%
%\begin{widetext}
%\begin{align}
\begin{align}
\langle0\mid J^{M_c}(x)\mid {M_c}(p, s)\rangle=&\lambda_{M_c} \gamma_5 \, u(p,s),\label{edmn04}\\
%\nonumber\\
\langle {M_c}(p+q, s)\mid\bar J^{M_c}(0)\mid 0\rangle=&\lambda_{M_c} \gamma_5 \, \bar u(p+q,s)\label{edmn004}
,\\
\langle {M_c}(p, s)\mid {M_c}(p+q, s)\rangle_\gamma &=\varepsilon^\mu\,\bar u(p, s)\bigg[\big[F_1(q^2)
+F_2(q^2)\big] \gamma_\mu +F_2(q^2)
\frac{(2p+q)_\mu}{2 m_{M_c}}\bigg]\,u(p+q, s), \label{edmn005}
%%%%%%%%%%%%%%%%%%%%%%%%%%%%%%%%%%
\end{align}
\begin{align}
\langle0\mid J_{\mu}^{M_c^*}(x)\mid {M_c^*}(p,s)\rangle&=\lambda_{{M_c^*}}u_{\mu}(p,s),\\
\langle {M_c^*}(p+q,s)\mid
\bar{J}_{\nu}^{M_c^*}(0)\mid 0\rangle &= \lambda_{{M_c^*}}\bar u_{\nu}(p+q,s), \\
%%%%%%%%%%%%%%%%%%%%%%%%%%%%%%%%%%%%%%%%%%%%%%%%%%%%%%%%%%%%%%%%%%%%%%%%%%%%
%%%%%%%%%%%%%%%%%%%%%%%%%%%%%%%%%%%%%%%%%%%%%%%%%%%%%%%%%%%%%%%%%%%%%%%%%%%%
\langle {M_c^*}(p,s)\mid {M_c^*}(p+q,s)\rangle_\gamma &=-e\bar
u_{\mu}(p,s)\bigg[F_{1}(q^2)g_{\mu\nu}\eslash 
-
\frac{1}{2m_{{M_c^*}}} 
\Big[F_{2}(q^2)g_{\mu\nu} 
+F_{4}(q^2)\frac{q_{\mu}q_{\nu}}{(2m_{{M_c^*}})^2}\Big]\eslash\qslash
\nonumber\\
&+
F_{3}(q^2)\frac{1}{(2m_{{M_c^*}})^2}q_{\mu}q_{\nu}\eslash \bigg] 
u_{\nu}(p+q,s),
\label{matelpar}
\end{align}
%\end{align}
%
where $F_i$'s are the Lorentz invariant form factors, the $u(p,s)$, $ u(p+q,s)$ and $\lambda_{{M_c}}$ are the spinors and residue of the $M_c$ state, respectively, and; the $u_{\mu}(p,s)$, $u_{\nu}(p+q,s)$ and $\lambda_{{M_c^*}}$ are the spinors and residue of the $M_c^*$ state, respectively.

Using the Eqs.~(\ref{edmn04})-(\ref{matelpar}) and some mathematical simplifications, we derive  the following results for the hadronic representation of the correlation functions,
\begin{align}
\label{edmn05}
\Pi^{Had}(p,q)=&\lambda^2_{M_c}\gamma_5 \frac{\Big(\pslash+m_{M_c} \Big)}{[p^{2}-m_{{M_c}}^{2}]}\varepsilon^\mu \bigg[\big[F_1(q^2) %
%\nonumber\\
%&
+F_2(q^2)\big] \gamma_\mu
+F_2(q^2)\, \frac{(2p+q)_\mu}{2 m_{M_c}}\bigg]  \gamma_5 
\frac{\Big(\pslash+\qslash+m_{M_c}\Big)}{[(p+q)^{2}-m_{{M_c}}^{2}]}, \\
\nonumber\\
%\end{align}
%\begin{align}
\Pi^{Had}_{\mu\nu}(p,q)&=\frac{\lambda_{_{{M_c^*}}}^{2}}{[(p+q)^{2}-m_{_{{M_c^*}}}^{2}][p^{2}-m_{_{{M_c^*}}}^{2}]} 
\bigg[  g_{\mu\nu}\pslash\eslash\qslash \,F_{1}(q^2) 
%\nonumber\\
%&
-m_{{M_c^*}}g_{\mu\nu}\eslash\qslash\,F_{2}(q^2)
-
\frac{F_{3}(q^2)}{4m_{{M_c^*}}}q_{\mu}q_{\nu}\eslash\qslash \nonumber\\
&
-
\frac{F_{4}(q^2)}{4m_{{M_c^*}}^3}(\varepsilon.p)q_{\mu}q_{\nu}\pslash\qslash 
+
\cdots %\mbox{other independent structures}
\bigg]. \label{final phenpart}
\end{align}
%
%To obtain the above expression, summing over the spins of $M_c$ state, $\sum_s u(p,s)\bar u(p,s)=\pslash+m_{M_c}$ and $\sum_s u(p+q,s)\bar u(p+q,s)=(\pslash+\qslash)+m_{M_c}$ have also been performed.

%

  The MDMs of hadrons are associated with their magnetic form factors; more specifically, the MDMs are equal to this form factor at zero momentum squared.  The magnetic form factors are defined by the form factors $F_i(q^2)$, which are more directly accessible in experiments and  are given by
\begin{align}
\label{edmn07}
F_M(q^2) &= F_1(q^2) + F_2(q^2),\\
G_{M}(q^2) &= \left[ F_1(q^2) + F_2(q^2)\right] ( 1+ \frac{4}{5}
\tau ) -\frac{2}{5} \left[ F_3(q^2)  \right]
+\left[
F_4(q^2)\right] \tau \left( 1 + \tau \right),
\end{align}  where $F_M(q^2)$ and $G_M(q^2)$ are magnetic form factor for spin-1/2 and spin-3/2 states, respectively and;  $\tau
= -\frac{q^2}{4m^2_{{M_c^*}}}$.  We can define the magnetic form factors for the MDMs $\mu_{M_c}$ and $\mu_{{M_c^*}}$ since we are dealing with a real photon, i.e. $q^2=0$:
\begin{align}
\label{edmn08}
\mu_{M_c} &= \frac{ e}{2\, m_{M_c}} \,F_M( 0),\\
\mu_{{M_c^*}}&=\frac{e}{2m_{{M_c^*}}}G_{M}(0).
\end{align}

Let's begin with the second step, which calculates the correlation functions utilizing QCD parameters. The second representation of the correlation function, the QCD side, is obtained by explicitly using the interpolating currents in the correlation function.  Following this, we perform Wick's theorem to contract all quark fields and obtain the desired results.  For example, in the case of the $D\Xi$ and  $D^*\Xi$ states, the results of the contractions are obtained as follows
\begin{align}
\label{QCD2}
\Pi^{QCD}(p,q)&=- \frac{i}{2}\varepsilon^{abc} \varepsilon^{a^{\prime}b^{\prime}c^{\prime}}\, \int d^4x \, e^{ip\cdot x} 
\nonumber\\
& 
 \langle 0\mid \Big\{ \mbox{Tr}\Big[\gamma_5 S_{c}^{dd^\prime}(x) \gamma_5  S_{u}^{d^\prime d}(-x)\Big]  
\mbox{Tr}\Big[\gamma_{\alpha} S_s^{bb^\prime}(x) \gamma_{\beta}  
  \widetilde S_{s}^{aa^\prime}(x)\Big]
(\gamma^{\alpha}\gamma_5 S_{u}^{cc^\prime}(x) \gamma_5  \gamma^{\beta})
 \nonumber\\
&     
- \mbox{Tr}\Big[\gamma_5 S_{c}^{dd^\prime}(x) \gamma_5  S_{u}^{d^\prime d}(-x)\Big]   
 \mbox{Tr}\Big[\gamma^{\alpha} S_{s}^{ba^\prime}(x)
   \gamma_{\beta} \widetilde S_s^{ab^\prime}(x)\Big] 
(\gamma^{\alpha}\gamma_5 S_{u}^{cc^\prime}(x) \gamma_5  \gamma^{\beta})\nonumber\\
&
  + \mbox{Tr}\Big[\gamma_5 S_{c}^{dd^\prime}(x) \gamma_5  S_{d}^{d^\prime d}(-x)\Big]  
\mbox{Tr}\Big[\gamma_{\alpha} S_s^{bb^\prime}(x) 
 \gamma_{\beta} \widetilde S_{s}^{aa^\prime}(x)\Big]
(\gamma^{\alpha}\gamma_5 S_{d}^{cc^\prime}(x) \gamma_5  \gamma^{\beta})
 \nonumber\\
&  -  \mbox{Tr}\Big[\gamma_5 S_{c}^{dd^\prime}(x) \gamma_5  S_{d}^{d^\prime d}(-x)\Big]   
 \mbox{Tr}\Big[\gamma^{\alpha} S_{s}^{ba^\prime}(x) 
 \gamma_{\beta} \widetilde S_s^{ab^\prime}(x) \Big] 
(\gamma^{\alpha}\gamma_5 S_{d}^{cc^\prime}(x) \gamma_5  \gamma^{\beta})
 \Big\}
\mid 0 \rangle _\gamma \,,\\
%\end{align}
\nonumber\\
%\begin{align}
\Pi_{\mu\nu}^{QCD}(p,q)&= - \frac{i}{2}\varepsilon^{abc} \varepsilon^{a^{\prime}b^{\prime}c^{\prime}}\, \int d^4x \, e^{ip\cdot x}   
\nonumber\\
& 
\langle 0\mid \Big\{ \mbox{Tr}\Big[\gamma_\mu S_{c}^{dd^\prime}(x) \gamma_\nu  S_{u}^{d^\prime d}(-x)\Big]  
\mbox{Tr}\Big[\gamma_{\alpha} S_s^{bb^\prime}(x) \gamma_{\beta}  \widetilde S_{s}^{aa^\prime}(x)\Big]
(\gamma^{\alpha}\gamma_5 S_{u}^{cc^\prime}(x) \gamma_5  \gamma^{\beta})
 \nonumber\\
&     
- \mbox{Tr}\Big[\gamma_\mu S_{c}^{dd^\prime}(x) \gamma_\nu  S_{u}^{d^\prime d}(-x)\Big]   
 \mbox{Tr}\Big[\gamma^{\alpha} S_{s}^{ba^\prime}(x)  \gamma_{\beta} \widetilde S_s^{ab^\prime}(x)\Big]
(\gamma^{\alpha}\gamma_5 S_{u}^{cc^\prime}(x) \gamma_5  \gamma^{\beta})\nonumber\\
&
  + \mbox{Tr}\Big[\gamma_\mu S_{c}^{dd^\prime}(x) \gamma_\nu  S_{d}^{d^\prime d}(-x)\Big]  
\mbox{Tr}\Big[\gamma_{\alpha} S_s^{bb^\prime}(x) 
 \gamma_{\beta} \widetilde S_{s}^{aa^\prime}(x)\Big]
(\gamma^{\alpha}\gamma_5 S_{d}^{cc^\prime}(x) \gamma_5  \gamma^{\beta})
 \nonumber\\
&  -  \mbox{Tr}\Big[\gamma_\mu S_{c}^{dd^\prime}(x) \gamma_\nu  S_{d}^{d^\prime d}(-x)\Big]   
 \mbox{Tr}\Big[\gamma^{\alpha} S_{s}^{ba^\prime}(x)  \gamma_{\beta} \widetilde S_s^{ab^\prime}(x) \Big] 
(\gamma^{\alpha}\gamma_5 S_{d}^{cc^\prime}(x) \gamma_5  \gamma^{\beta})
 \Big\}
\mid 0 \rangle _\gamma \,,\label{QCD4}
\end{align}
where   
%\begin{equation*}
$\widetilde{S}_{c(q)}^{ij}(x)=CS_{c(q)}^{ij\mathrm{T}}(x)C$. The full propagators for both heavy and light quarks are denoted as $S_{c}(x)$ and $S_{q}(x)$, respectively, and can be expressed in the following way:~\cite{Yang:1993bp, Belyaev:1985wza}
\begin{align}
\label{edmn13}
S_{q}(x)&= S_q^{free}(x) 
- \frac{\langle \bar qq \rangle }{12} \Big(1-i\frac{m_{q} \xslash}{4}   \Big)
%\nonumber\\
%&
- \frac{ \langle \bar qq \rangle }{192}
m_0^2 x^2  \Big(1
  -i\frac{m_{q} \xslash}{6}   \Big)
-\frac {i g_s }{32 \pi^2 x^2} ~G^{\mu \nu} (x) 
%\nonumber\\
%& \times 
\Big[\rlap/{x} 
\sigma_{\mu \nu} +  \sigma_{\mu \nu} \rlap/{x}
 \Big],\\
%\nonumber\\
%\end{align}%
%and
%%
%\begin{align}
S_{c}(x)&=S_c^{free}(x)
%\nonumber\\
%&
-\frac{g_{s}m_{c}}{16\pi ^{2}} \int_0^1 dv\, G^{\mu \nu }(vx)\Bigg[ (\sigma _{\mu \nu }{\xslash}
  +{\xslash}\sigma _{\mu \nu }) 
    \frac{K_{1}\Big( m_{c}\sqrt{-x^{2}}\Big) }{\sqrt{-x^{2}}}
   %\nonumber\\
  %&
 +2\sigma_{\mu \nu }K_{0}\Big( m_{c}\sqrt{-x^{2}}\Big)\Bigg],
 \label{edmn14}
\end{align}%
and 
\begin{align}
 S_q^{free}(x)&=\frac{1}{2 \pi x^2}\Big(i \frac{\xslash}{x^2}- \frac{m_q}{2}\Big),\\
 S_c^{free}(x)&=\frac{m_{c}^{2}}{4 \pi^{2}} \Bigg[ \frac{K_{1}\Big(m_{c}\sqrt{-x^{2}}\Big) }{\sqrt{-x^{2}}}
+i\frac{{\xslash}~K_{2}\Big( m_{c}\sqrt{-x^{2}}\Big)}
{(\sqrt{-x^{2}})^{2}}\Bigg],
\end{align}
where $\langle \bar qq \rangle$ represents the light-quark condensate, $G^{\mu\nu}$ corresponds to the gluon field strength tensor, $v$ stands for line variable, and $K_i$'s refer to the modified Bessel functions of the second kind, respectively.     

The correlation functions in Eqs.~(\ref{QCD2}) and (\ref{QCD4}) receive both perturbative contributions, that is, when a photon interacts with light/heavy quarks at short-distance and non-perturbative contributions, that is when a photon interacts with light quarks at large-distance. 
In the case of perturbative contributions, one of the light propagators or one of the heavy propagators of the quarks interacting perturbatively with the photon is replaced by the following
\begin{align}
\label{free}
S^{free}(x) \rightarrow \int d^4y\, S^{free} (x-z)\,\rlap/{\!A}(z)\, S^{free} (z)\,,
\end{align}
and the rest of the propagators in Eqs.~(\ref{QCD2}) and (\ref{QCD4}) are considered full quark propagators. Here we use $ A_\mu(z)=\frac{i}{2}z^{\nu}(\varepsilon_\mu q_\nu-\varepsilon_\nu q_\mu)\,e^{iq.z} $.

 For the calculation of the non-perturbative contributions, one of the light-quark propagators in Eq.~(\ref{edmn13}), which characterizes the photon emission at large distances, is replaced by
\begin{align}
\label{edmn21}
S_{\alpha\beta}^{ab}(x) \rightarrow -\frac{1}{4} \Big[\bar{q}^a(x) \Gamma_i q^b(0)\Big]\Big(\Gamma_i\Big)_{\alpha\beta},
\end{align}
and in Eqs.~(\ref{QCD2}) and (\ref{QCD4}) the remaining four quark propagators are all considered to be full quark propagators. 
Here $\Gamma_i = \{\textbf{1}, \gamma_5, \gamma_\mu, i\gamma_5 \gamma_\mu, \sigma_{\mu\nu}/2\}$. Once Eq.~(\ref{edmn21}) is inserted into Eqs.~(\ref{QCD2}) and (\ref{QCD4}), matrix elements of type $\langle \gamma(q)\vel \bar{q}(x) \Gamma_i q(0) \ver 0\rangle$ and $\langle \gamma(q)\vel \bar{q}(x) \Gamma_i G_{\alpha\beta}q(0) \ver 0\rangle$ appear which are parameterized in terms of photon DAs and describes the interaction of photons with quark fields at large distance.  These matrix elements, which are parameterized by photon wave functions with specific twists, are essential for computing the non-perturbative contributions (for explicit expressions of photon distribution amplitudes (DAs), refer to Ref.~\cite{Ball:2002ps}).  
The correlation functions related to the quark-gluon degrees of freedom and DAs of the photon can be attained through the use of Eqs.~(\ref{QCD2})$-$(\ref{edmn21}) via QCD representation.

The QCD light cone sum rules are derived by matching the expressions of the correlation function using the QCD parameters and those using the hadron properties, by means of the spectral representation. To eliminate the effects that come from the higher states and the continuum, we utilize both the Borel transformation and the continuum subtraction provided by the quark-hadron duality approximation.  The results of performing all of the aforementioned procedures on the MDMs are as follows:
\begin{align}
\label{edmn15}
\mu_{D\Xi}  &=\frac{e^{\frac{m^2_{D\Xi}}{\mathrm{M^2}}}}{\,\lambda^2_{D\Xi}\, m_{D\Xi}}\, \Delta_1^{QCD} (\mathrm{M^2},\mathrm{\mathrm{s_0}}),\\
\mu_{D^*\Xi} & =\frac{e^{\frac{m^2_{D^*\Xi}}{\mathrm{M^2}}}}{\,\lambda^2_{D^*\Xi}\, m_{D^*\Xi}}\, \Delta_2^{QCD} (\mathrm{M^2},\mathrm{\mathrm{s_0}}).
\end{align}
For the sake of simplicity, we provide the explicit form of the $\Delta_1^{QCD}(\mathrm{M^2},\mathrm{s_0})$ function as an example in the Appendix. As shown in the equation above, there are two additional parameters: the continuum threshold $\mathrm{s_0}$ and the Borel parameters $\mathrm{\mathrm{M^2}}$. In the section devoted to numerical analysis, we will go into detail about the process of determining the working intervals for these additional parameters.  

Analytical results have been obtained for the singly-charmed molecular pentaquark states. Next, numerical calculations will be conducted for these states.  

\end{widetext}

\section{Numerical illustrations } \label{secIII}

This section presents the numerical analysis for the MDMs of singly-charmed molecular pentaquark states.  The following parameters are used in our calculations:
  $m_u=m_d=0$, $m_s =93.4^{+8.6}_{-3.4}\,\mbox{MeV}$,
$m_c = 1.27\pm 0.02\,$GeV~\cite{ParticleDataGroup:2022pth},  
$m_{\Omega_c(3185)} = 3185.1\pm 1.7^{+7.4}_{-0.9}\pm0.2 \mbox{ MeV}$, 
$ m_{\Omega_c(3327)} = 3327.1\pm1.2^{+0.1}_{-1.3} \pm 0.2\mbox{ MeV}$~\cite{LHCb:2023sxp},  
$m_{D_s \Xi} =  3.28^{+0.12}_{-0.12}$~GeV, 
$m_{D_s^* \Xi} =  3.44^{+0.11}_{-0.11}$~GeV,  
$m_{D \Xi^*} =  3.41^{+0.11}_{-0.12}$~GeV, 
$m_{D_s \Xi^*} =  3.51^{+0.11}_{-0.13}$~GeV,  
$\lambda_{D \Xi} =( 1.42^{+0.27}_{-0.24}) \times 10^{-3}$\,GeV$^6$, 
$\lambda_{D_s \Xi} =( 1.60^{+0.30}_{-0.27}) \times 10^{-3}$\,GeV$^6$, 
$\lambda_{D^* \Xi} =( 1.59^{+0.29}_{-0.26}) \times 10^{-3}$\,GeV$^6$, 
$\lambda_{D_s^* \Xi} =( 1.72^{+0.30}_{-0.27}) \times 10^{-3}$\,GeV$^6$, 
$\lambda_{D \Xi^*} =( 1.07^{+0.19}_{-0.17}) \times 10^{-3}$\,GeV$^6$, 
$\lambda_{D_s \Xi^*} =( 1.12^{+0.21}_{-0.19}) \times 10^{-3}$\,GeV$^6$ 
~\cite{Xin:2023gkf}, $\langle \bar ss\rangle $= $0.8 \langle \bar qq\rangle$ with $\langle \bar qq\rangle $=$(-0.24\pm 0.01)^3\,$GeV$^3$~\cite{Ioffe:2005ym},   
$m_0^{2} = 0.8 \pm 0.1$~GeV$^2$, $\langle g_s^2G^2\rangle = 0.88~ $GeV$^4$~\cite{Matheus:2006xi}, $\chi=-2.85 \pm 0.5~\mbox{GeV}^{-2}$~\cite{Rohrwild:2007yt} and $f_{3\gamma}=-0.0039~$GeV$^2$~\cite{Ball:2002ps}.  The photon DAs and the parameters used in those DAs are taken from Ref.~\cite{Ball:2002ps}.

Besides the above-mentioned parameters, the sum rules for the MDMs also depend on the helping parameters: the Borel mass squared parameter $\mathrm{M^2}$ and the continuum threshold $\mathrm{s_0}$.   Physical measurables, in our case MDMs, should be varied as slightly as possible by these helping parameters.  Hence, we need to find the so-called working windows of these helping parameters, where physical observables, in our case MDMs, are insensitive to the variation of these parameters in their working windows. The standard prescription of the technique used, the operator product expansion ($\mathrm{OPE}$) convergence and the pole contribution ($\mathrm{PC}$) dominance are taken into account in the determination of the working ranges for the parameters $\mathrm{M^2}$ and $\mathrm{s_0}$. For the purposes of defining the aforementioned restrictions, it is useful to utilize the formulas presented below:
\begin{align}
 \mathrm{PC}&=\frac{\Delta (\mathrm{M^2},\mathrm{s_0})}{\Delta (\mathrm{M^2},\infty)} \geq  30\%,\\
%%%%%%%%%%%%%%%%%%%%%%%%%%
 \mathrm{OPE} &=\frac{\Delta^{\scriptsize{\mbox{DimN}}} (\mathrm{M^2},\mathrm{s_0})}{\Delta (\mathrm{M^2},\mathrm{s_0})}\leq  5\%,
 \end{align}
 where $\Delta^{\mathrm{DimN}}(\mathrm{M^2},\mathrm{s_0})=\Delta^{\mathrm{Dim(10+11+12)}}(\mathrm{M^2},\mathrm{s_0})$.   
The working windows for the helping parameters are presented in Table \ref{parameter} on the basis of these restrictions. It follows that the working windows determined for $\mathrm{M^2}$ and $\mathrm{s_0}$ fulfill the constraints imposed by the dominance of $\mathrm{PC}$ and the convergence of $\mathrm{OPE}$.  Having determined the working windows of $\mathrm{\mathrm{M^2}}$ and $\mathrm{s_0}$, we can explore the variation of the MDMs on $\mathrm{M^2}$ for different values of $\mathrm{s_0}$.  For example, in Figure 1, the MDMs of the $D \Xi$ and $D^* \Xi$ states are plotted, and it is seen that the MDMs of these states show good stability regarding the variation of $\mathrm{M^2}$ in its working interval.
\begin{widetext}
 
 \begin{table}[htp]
	\addtolength{\tabcolsep}{10pt}
	\caption{Working windows of  $\mathrm{s_0}$ and  $\mathrm{M^2}$ as well as the PC and OPE convergence for the MDMs of singly-charmed molecular pentaquark states.}
	\label{parameter}
		\begin{center}
		%\scalebox{1.0}{
\begin{tabular}{l|ccccc}
                \hline\hline
                \\
State & ~~$\mathrm{s_0}$ (GeV$^2$)~~& ~~$\mathrm{M^2}$ (GeV$^2$) ~~& ~~  $\mathrm{PC}$ ($\%$) ~~ & ~~  $\mathrm{OPE}$  
 ($\%$) \\
 \\
                                        \hline\hline
                                        \\
$D \Xi$  & $13.5-15.5$ & $3.5-5.5$ & $35-55$ &  $3.82$  
                        \\
 %                       \\
$ D_s \Xi$ & $14.0-16.0$ & $3.5-5.5$ & $36-56$ &  $3.91$  \\
                       \\
                                        \hline\hline
                                        \\
$ D^* \Xi$ & $14.5-16.5$ & $3.5-5.5$ & $36-55$ &  $3.85$  \\
 %                       \\
$ D_s^* \Xi$ & $15.5-17.5$ & $3.7-5.7$ & $36-58$ &  $3.76$   \\
  %                      \\
$ D \Xi^*$   & $15.0-17.0$ &$3.7-5.7$ & $35-57$ & $3.83$   \\
   %                      \\
$D_s \Xi^*$ & $16.0-18.0$ & $3.7-5.7$ & $36-58$ &  $3.73$   \\
                       \\
                                       \hline\hline
 \end{tabular}
%}
\end{center}
\end{table}

\end{widetext}

\begin{widetext}

\begin{figure}[htp]
\label{Msqfig}
\centering
 \includegraphics[width=0.45\textwidth]{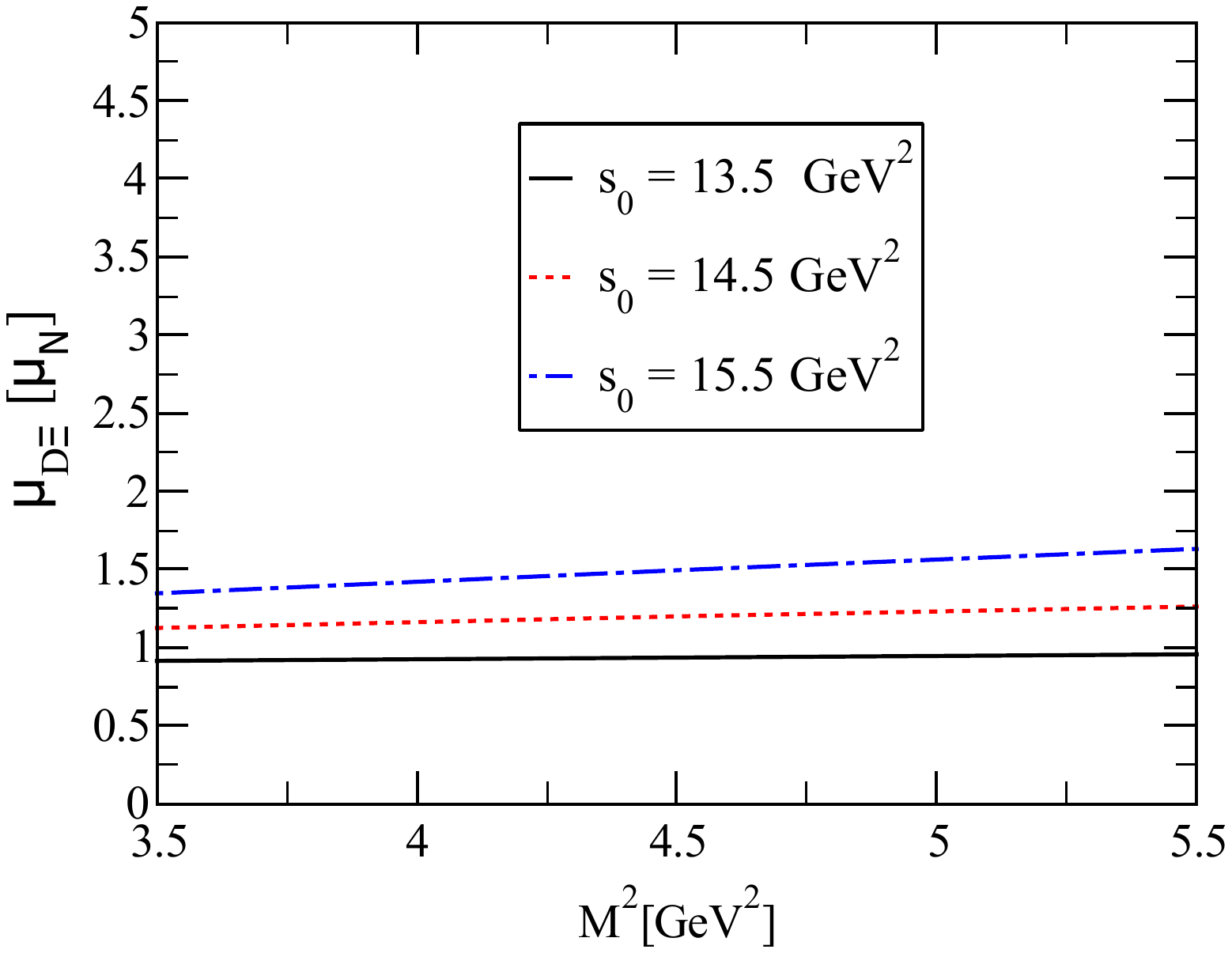}~~%\\
 %\vspace{0.5 cm}%~~~ 
 \includegraphics[width=0.45\textwidth]{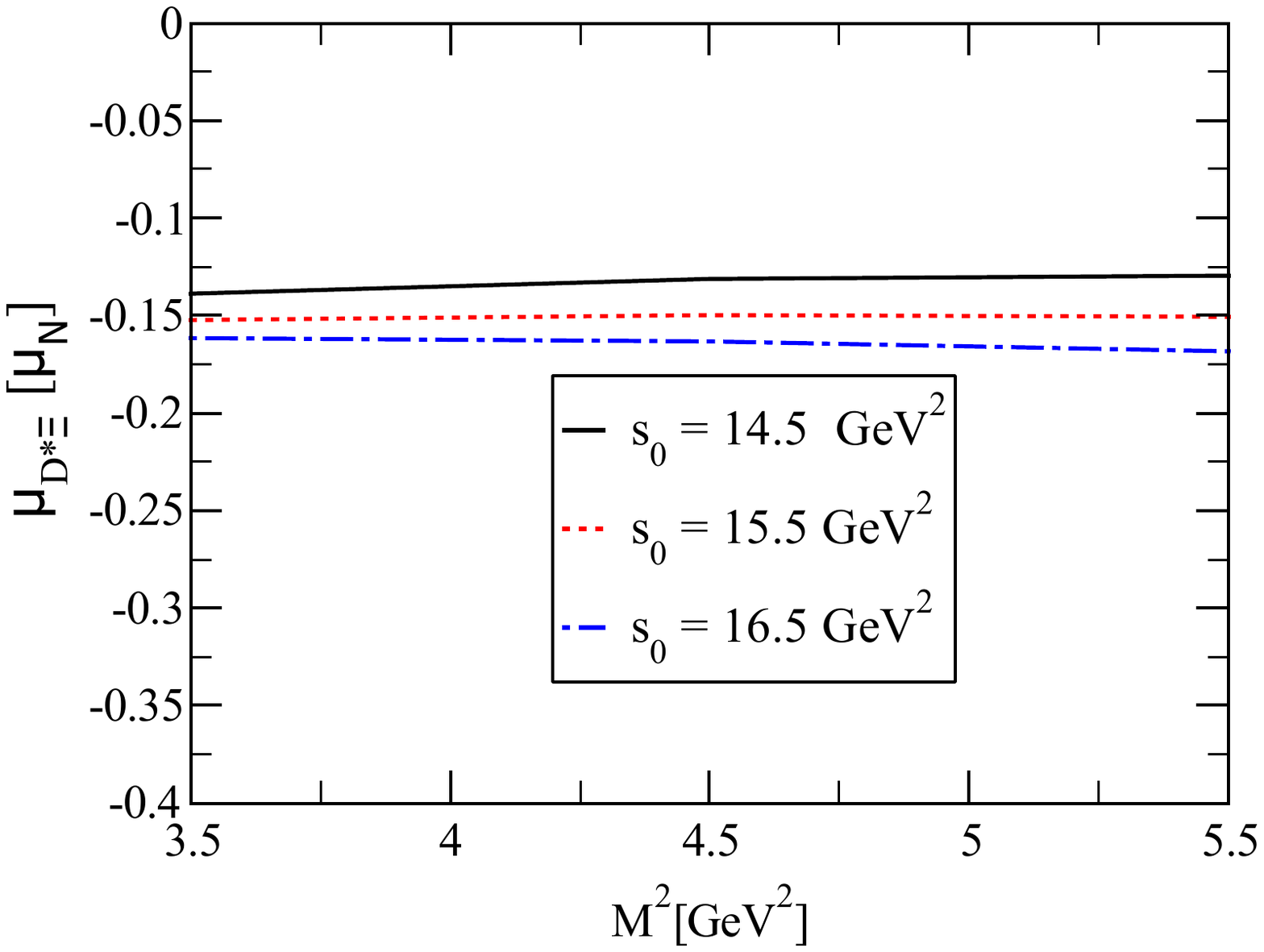}
 \caption{Dependencies of the MDMs of $D\Xi$ and $D^*\Xi$ states on $\mathrm{M^2}$ at three values of $\mathrm{s_0}$.}
  \end{figure}
 
  \end{widetext}
  %%%%%%%%%%%%%%%%%%%%%%%%%%%%%%%%%%%%%%%%%%%%%

To give numerical values for the MDMs of the singly-charmed molecular pentaquark states, we have determined all the necessary parameters. As a result of our extensive numerical computations, the obtained MDMs are shown in Table \ref{sonuc}.  We would like to point out that the numerical calculations take into account the uncertainty of the input parameters, the uncertainties of the numerical parameters entering the DAs of the photon, and the uncertainty caused by the variation of the helping parameters $\mathrm{M^2}$ and $\mathrm{s_0}$. It follows from these results that the MDMs of these states are sufficiently large to be measured in future experiments. Measurements of these MDMs will be of great help in understanding the structure and nature of these controversial states. Our results can also be checked by other non-perturbative approaches.
 \begin{widetext}

  \begin{table}[htp]
	\addtolength{\tabcolsep}{10pt}
	\caption{The MDMs of the $\Omega_{c}(3185)^0$,  $\Omega_c(3327)^0$ and other possible singly-charmed molecular pentaquark states derived by running the QCD light-cone sum rules.}
	\label{sonuc}
		\begin{center}
		%\scalebox{1.0}{
\begin{tabular}{l|ccccccc}
                \hline\hline
               \\
 Parameters& $ D \Xi$ & $ D_s \Xi^0$ & $D_s \Xi^-$  & - & -&- \\
    \\
                                        \hline\hline
                                      \\

 $\mu [\mu_N]$& $1.27 \pm 0.35 $ & $1.05 \pm 0.29 $ & $ 0.50 \pm 0.10 $ & -  & -&-                        \\
                         \\
                                        \hline\hline
      \\  
       Parameters & $D^* \Xi$& $D_s^* \Xi^0$ &$D_s^* \Xi^- $  & $D \Xi^*$ & $D_s \Xi^{*0}$& $D_s \Xi^{*-}$ \\
      \\
 \hline \hline
 \\
$\mu [\mu_N]$ & $-0.15 \pm 0.02$ & $-0.12 \pm 0.02$ & $0.57 \pm 0.06 $ &  $0.23 \pm 0.05$  & $1.04 \pm 0.21$& $-0.54 \pm 0.11$ \\
                        \\
                                       \hline\hline
 \end{tabular}
%}
\end{center}
\end{table}

 \end{widetext}

\section{summary and concluding remarks}\label{secIV}

In summary, assuming that $\Omega_{c}(3185)^0$ and $\Omega_c(3327)^0$ are meson-baryon configurations and using the photon DAs, the MDMs of these states have been extracted within the QCD light-cone sum rules. Along with these states, we have also obtained the MDMs of possible $D_s \Xi$, $D_s^* \Xi$, $D \Xi^*$, and $D_s \Xi^*$ pentaquark states.  The obtained result may be useful in determining the exact nature of these states.  To understand the inner structure of the $\Omega_{c}(3185)^0$ and $\Omega_c(3327)^0$ states, the confirmation of our predictions with other non-perturbative approaches and future experiments can be very helpful.
As an example, the MDM of the $\Lambda(1405)$ state was calculated using the lattice QCD and assuming that the $\Lambda(1405)$ state is a $\bar K N$  meson-baryon configuration~\cite{Hall:2014uca,Hall:2016kou}. 
 We therefore recommend a comparable investigation of the MDMs of the $\Omega_{c}(3185)^0$ and $\Omega_c(3327)^0$ states using lattice QCD. 
 Compared to the estimated MDMs of lattice QCD, the predictions of this study can yield further insight for the experimental search for the $\Omega_{c}(3185)^0$,  $\Omega_c(3327)^0$, and other possible states, and the experimental measurements for these MDMs can be a crucial test for the meson-baryon nature of these states.  
Experimental facilities like LHCb, BESIII, Belle II,  etc. are capable of measuring MDMs of $\Omega_{c}(3185)^0$, $\Omega_c(3327)^0$, and other possible states with the increased luminosity in future runs and our predictions.
We hope that our predictions of the MDMs of the $\Omega_{c}(3185)^0$, $\Omega_c(3327)^0$, and other possible states, along with the results of other theoretical studies of the mass and decay widths of these intriguing states, can be very useful in searching for them in future experiments and in defining the exact nature of these states.

  \begin{widetext}

%\appendix
\section*{Appendix: Explicit forms of the $\Delta_1^{QCD}(\mathrm{M^2},\mathrm{s_0})$ function } \label{appenda}
In this appendix,  the explicit forms of the expressions achieved for the MDM of the $\Omega_{c}(3185)^0$ state are given as follows: 
\begin{align}
 \Delta_1^{QCD}(\mathrm{M^2},\mathrm{\mathrm{s_0}})&= \frac {P_ 1 P_ 2^2 (e_d + 
     3 e_u) } {110592 m_c \pi^4}\Bigg[- (m_c^4 I[-2, 0] - 
       I[0, 0])(A[u_ 0] - 
        I_ 3[h_ {\gamma}]) + \chi \Big (4 m_c^6 I[-3, 1] + 
        m_c^4 \big (-m_ 0^2 I[-2, 0] \nonumber\\
        &+ 2 I[-2, 1]\big) + 
        m_ 0^2 I[0, 0] - 2 I[0, 1]\Big) \varphi_ {\gamma}[u_ 0]\Bigg]\nonumber%\\
        \end{align}
\begin{align}
        %%%%%%%%%%%%%%%%%%%%%%%%%%%%%%%%%%%%%%%%%%%%%%%%%%%%%%%%%%%%%%%%%%%%%%%%%%
        &+\frac {m_s P_ 1 P_ 2 P_ 3} {442368 m_c^2 \pi^4} \Bigg[
   48 (e_d + 3 e_u)  (m_c^4 I[-2, 0] - I[0, 0]) A[u_ 0] - 
    13 (e_d + 3 e_u)  (m_c^4 I[-2, 0] - I[0, 0]) I_ 2[\mathcal S]\nonumber\\
    &- 
    48 (e_d - 8 e_s + 3 e_u)  (m_c^4 I[-2, 0] - I[0, 0]) I_ 3[
      h_ {\gamma}] + 384 e_s (-m_c^4 I[-2, 0] + I[0, 0]) + 
    32 \chi \Bigg (e_d \Big (-6 m_c^6 I[-3, 1]
      \nonumber\\
    & + 
           m_c^4 (m_ 0^2 I[-2, 0] - 3 I[-2, 1]) - m_ 0^2 I[0, 0] + 
           3 I[0, 1]\Big) + 
        3 \Big (8 e_s (m_c^4 I[-2, 1]+ I[0, 1]) + 
            e_u (-6 m_c^6 I[-3, 1] \nonumber\\
    & + 
                m_c^4 (m_ 0^2 I[-2, 0] - 3 I[-2, 1]) - 
                m_ 0^2 I[0, 0] + 
                3 I[0, 1])\Big)\Bigg) \varphi_ {\gamma}[u_ 0]\Bigg]\nonumber\\
                %%%%%%%%%%%%%%%%%%%%%%%%%%%%%%%%%%%%%%%%%%%%%%%%%%%%%%%%%%%%%%%%%%%%%%%%%%%%%
                &+\frac {m_s P_ 2^2 P_ 3 } {2304 m_c \pi^2}\Bigg[
   96 e_s (m_c^4 I[-2, 0] - I[0, 0]) - 
    12 (e_d + 3 e_u)  (m_c^4 I[-2, 0] - I[0, 0]) A[u_ 0] + 
    3 (e_d + 3 e_u)  (m_c^4 I[-2, 0]  \nonumber\\
    &- I[0, 0]) I_ 2[\mathcal S] + 
    12 (e_d - 8 e_s + 3 e_u) (m_c^4 I[-2, 0] - I[0, 0]) I_ 3[
      h_ {\gamma}] - 
    4 \chi \Bigg (48 e_s (m_c^4 I[-2, 1]+ I[0, 1]) + 
        e_d (-12 m_c^6 I[-3, 1] \nonumber\\
    & + 
           m_c^4 (5 m_ 0^2 I[-2, 0] - 6 I[-2, 1]) - 
           5 m_ 0^2 I[0, 0] + 6 I[0, 1]) 
           + 
        3 e_u \Big (-12 m_c^6 I[-3, 1] + 
            m_c^4 (5 m_ 0^2 I[-2, 0] - 6 I[-2, 1])
           \nonumber\\
    & - 
            5 m_ 0^2 I[0, 0] + 6 I[0, 1]\Big)\Bigg) \varphi_ {\gamma}[
       u_ 0]\Bigg] \nonumber\\
       %%%%%%%%%%%%%%%%%%%%%%%%%%%%%%%%%%%%%%%%%%%%%%%%%%%%%%%%%%%%%%%%%%%%%%%%%%%%%%%%%%%%%%%%%%%%%%
       &-\frac {e_s P_ 2 P_ 3^2} {24 m_c^2 \pi^2}\Bigg[(m_c^6 I[-3, 1] + 
       I[0, 1]) I_ 3[
      h_ {\gamma}] + \chi (m_c^6 I[-3, 2] - 
        I[0, 2]) \varphi_ {\gamma}[u_ 0]\Bigg]\nonumber\\
        %%%%%%%%%%%%%%%%%%%%%%%%%%%%%%%%%%%%%%%%%%%%%%%%%%%%5
        &+\frac {P_ 1 P_ 2} {3538944 m_c^2 \pi^6} \Bigg[-39 (e_d + 
       3 e_u) m_c^4  (m_c^4 I[-4, 2] - I[-2, 2]) I_ 2[\mathcal S] - 
    12 m_c^4(e_d + 3 e_u)   \big (7 m_c^4 I[-4, 2]- 
       13 m_c^2 I[-3, 2]
    \nonumber\\
    & + 6 I[-2, 2]\big) I_ 3[h_ {\gamma}] + 
    6 (e_d + 3 e_u)  \big (14 m_c^8 I[-4, 2] - 13 m_c^6 I[-3, 2] - 
       I[0, 2]\big) A[u_ 0] + 
    8 \Bigg (-39 e_s f_ {3\gamma} \pi^2 I_ 1[\mathcal V] (m_c^4 I[-2, 
             1]\nonumber\\
       & + I[0, 1]) + 
        144 e_s \Big (m_c^8 (I[-4, 2] - 2 I[-3, 1]) + 
           m_c^6 \big (m_ 0^2 (2 I[-3, 1] 
           + I[-2, 0]) - 4 I[-3, 2] - 
              I[-2, 1]\big)
              + 
           m_c^4 \big (3 I[-2, 2] 
           \nonumber\\
                   %%%%%%%%%%%%%%%%%%%%%%%%%%%%%%%%%%%%%%%%%%%%%%%%%%%%%%%%%%%%%%%%%%%%%%%%%%%%%%%%%%%%%%%%%%%%
    & + m_ 0^2 (I[-2, 1] - I[-1, 0]) - 
              2 I[-1, 1]\big) + 
           m_ 0^2 I[0, 1]\Big) 
           + \chi (e_d + 
            3 e_u) \Big (5 m_c^{10} I[-5, 3] + 7 m_c^8 I[-4, 3]
            - 
            2 m_c^4 I[-2, 3]
           \nonumber\\
    &  + 10 I[0, 3]\Big) \varphi_ {\gamma}[
           u_ 0]\Bigg)\Bigg]\nonumber\\
           %%%%%%%%%%%%%%%%%%%%%%%%%%%%%%%%%%%%%%%%%%%%%%%%%%%%%%%%%%%%%%%%%%%%%%%%%%%%%%%%%%%%%%%%%%
    &+\frac {P_ 2^2} {18432 m_c \pi^4}\Bigg[
   18 (e_d + 3 e_u) m_c^4 \Big (m_c^4 I[-4, 2] - 
       m_c^2 (m_ 0^2 I[-3, 1] + 2 I[-3, 2]) - m_ 0^2 I[-2, 1] 
       + 
       I[-2, 2]\Big) I_ 3[h_ {\gamma}]
       \nonumber\\
    & - 
    9 (e_d + 3 e_u)  \Big (2 m_c^8 I[-4, 2] - 
       2 m_c^6 (m_ 0^2 I[-3, 1] + I[-3, 2]) - m_ 0^2 m_c^4 I[-2, 1] 
       + 
       m_ 0^2 I[0, 1]\Big) A[u_ 0] + 
    9 (e_d + 3 e_u)
       \nonumber\\
    & \times \Big (m_c^8 I[-4, 2] - m_ 0^2 m_c^6 I[-3, 1] - 
       m_c^4 I[-2, 2] + m_ 0^2 I[0, 1]\Big) I_ 2[\mathcal S] 
       + 
    72 e_s \Bigg (-4 m_c^8 (I[-4, 2] - 2 I[-3, 1])
       \nonumber\\
    & - 
       8 m_c^6 \big (m_ 0^2 (2 I[-3, 1] + I[-2, 0]) - 
          2 (I[-3, 2] + I[-2, 1])\big)
          + 
       m_c^4 \Big (m_ 0^4 I[-2, 0] - 12 I[-2, 2] + 
          8 m_ 0^2 (-I[-2, 1]
          \nonumber\\
    & + I[-1, 0]) + 8 I[-1, 1]\Big) + 
       f_ {3\gamma} \pi^2  (m_c^4 I[-2, 1]
       + 
          I[0, 1]) I_ 1[\mathcal V] - 
       m_ 0^2 (m_ 0^2 I[0, 0] + 8 I[0, 1])\Bigg)
       \nonumber\\
            %   \end{align}
             %  \begin{align}
               & + 
    2 \chi (e_d + 
        3 e_u) \Big (m_c^8 \big (4 m_c^2 I[-5, 3] - 
           9 m_ 0^2 I[-4, 2] 
     %      \nonumber\\
    %&
    + 6 I[-4, 3]\big) + 
        9 m_ 0^2 m_c^6 I[-3, 2] - 2 m_c^4 I[-2, 3] + 
        8 I[0, 3]\Big) \varphi_ {\gamma}[u_ 0]\Bigg]\nonumber\\
        %%%%%%%%%%%%%%%%%%%%%%%%%%%%%%%%%%%%%%%%%%%%%%%%%%%%%%%%%%%%%%
                        & + \frac {m_s P_ 2 P_ 3} {6144 m_c^2 \pi^4} + \Bigg[(e_d + 
        3 e_u)  \Big (3 m_c^8 I[-4, 2] + m_ 0^2 m_c^6 I[-3, 1] + 
       m_ 0^2 I[0, 1] - 3 I[0, 2]\Big) I_ 2[\mathcal S] + 
    2 \bigg (m_c^6 \Big (-2 (e_d + 3 e_u) \nonumber\\
    & \times (3 m_c^2 I[-4, 2] + 
             m_ 0^2 I[-3, 1]) + 
          3 (e_d - 8 e_s + 3 e_u) I[-3, 2]\Big) - 
       2 (e_d + 3 e_u) m_ 0^2 I[0, 1] + 
       3 (e_d + 8 e_s + 3 e_u) I[0, 2]\bigg) A[u_ 0] 
       \nonumber%\\
\end{align}

\begin{align}
    &+ 
    4 \Bigg (-24 e_s m_c^8 I[-4, 
          2] + \bigg (m_c^6 \Big (3 e_d m_c^2 I[-4, 2] - 
              24 e_s m_c^2 I[-4, 2] + 9 e_u m_c^2 I[-4, 2] + 
              e_d m_ 0^2 I[-3, 1] - 12 e_s m_ 0^2 I[-3, 1] 
              \nonumber\\
    &+ 
              3 e_u m_ 0^2 I[-3, 1] - 
              3 (e_d - 8 e_s + 3 e_u) I[-3, 2]\Big) + (e_d + 
              3 (-4 e_s + e_u)) m_ 0^2 I[0, 1]\bigg) I_ 3[
          h_ {\gamma}] + 
        4 e_s \Big (-12 m_c^6 (m_c^2 I[-3, 1] 
        \nonumber\\
    &- I[-3, 2] + 
              I[-2, 1]) + 18 I[0, 2] + 
           5 m_ 0^2 (m_c^2 I[0, 0] + 2 I[0, 1] + 
               m_c^6 (2 I[-3, 1] - 
                  I[2, 0]))\Big) + \chi \bigg (e_d \Big (-2 m_c^{10} \
I[-5, 3] \nonumber\\
    &- 2 m_c^8 (m_ 0^2 I[-4, 2] + I[-4, 3]) + 
              m_ 0^2 m_c^6 I[-3, 2] + m_ 0^2 I[0, 2] - 
              4 I[0, 3]\Big) + 
           3 e_u \Big (-2 m_c^{10} I[-5, 3] - 
              2 m_c^8 (m_ 0^2 I[-4, 2] 
              \nonumber\\
    &+ I[-4, 3]) + 
              m_ 0^2 m_c^6 I[-3, 2] + m_ 0^2 I[0, 2] - 
              4 I[0, 3]\Big) + 
           4 e_s \Big (4 m_c^8 I[-4, 3] + 
               m_c^6 (-3 m_ 0^2 I[-3, 2] + 4 I[-3, 3]) + 
               3 m_ 0^2 I[0, 2]
               \nonumber\\
    &+ 
               8 I[0, 3]\Big)\bigg) \varphi_ {\gamma}[u_ 0]\Bigg)\Bigg]\nonumber\\
                              %%%%%%%%%%%%%%%%%%%%%%%%%%%%%%%%%%%%%%%%%%%%%%%%%%%%%%%%%%%%%%%%%%%%%%%%%%%%%%%%%%%%%%%%%%%%%%%%
               & + \frac {P_ 2 } {491520 m_c^2 \pi^6}\Bigg[
    10 (e_d + 3 e_u) m_c^6  \Big (m_c^6 I[-6, 4] - 3 m_c^4 I[-5, 4] + 
        3 m_c^2 I[-4, 4] - I[-3, 4]\Big) I_ 3[h_ {\gamma}] - 
     5 (e_d + 3 e_u) m_c^6  \nonumber\\
    & \times \Big (2 m_c^6 I[-6, 4] - 
        3 m_c^4 I[-5, 4] + I[-3, 4]\Big) A[u_ 0] + 
     5 (e_d + 3 e_u) m_c^6 \Big (m_c^6 I[-6, 4] - 3 m_c^2 I[-4, 4] + 
        2 I[-3, 4]\Big) I_ 2[\mathcal S] 
        \nonumber\\
    &- 
     60 e_s f_ {3\gamma} \pi^2  \Big (4 m_c^8 I[-4, 3] + 
        m_c^6 (-3 m_ 0^2 I[-3, 2] + 4 I[-3, 3]) + 3 m_ 0^2 I[0, 2] + 
        8 I[0, 3]\Big) I_ 1[\mathcal V] - 
     160 e_s \Bigg (m_c^{12} (3 I[-6, 4] 
     \nonumber\\
              % \end{align}
              % \begin{align}
    &+ 4 I[-5, 3]) + 
        3 m_c^{10} \Big (-4 I[-5, 4] + 
           m_ 0^2 (-4 I[-5, 3] + 3 I[-4, 2]) + 4 I[-4, 3]\Big) - 
        3 m_c^8 \Big (-5 I[-4, 4] + 6 m_ 0^2 (I[-4, 3]
        \nonumber\\
    &+ I[-3, 2]) - 
           4 I[-3, 3]\Big) + 
        m_c^6 \Big (-6 I[-3, 4] + m_ 0^2 (-6 I[-3, 3] + 9 I[-2, 2]) + 
           4 I[-2, 3]\Big) - 36 m_ 0^2 I[0, 3] + 
        32 m_c^2 I[0, 3]\Bigg) \nonumber\\
    &+ 
     4 \chi (e_d + 
         3 e_u) \bigg (m_c^6 \Big (m_c^6 I[-6, 5] + 
            3 m_c^4 I[-5, 5] + 3 m_c^2 I[-4, 5] + I[-3, 5]\Big) + 
         8 I[0, 5]\Bigg) \varphi_ {\gamma}[u_ 0]\Bigg],
               \end{align}
where $P_1 =\langle g_s^2 G^2\rangle$,  $P_2 =\langle \bar q q \rangle$ and  $P_3 =\langle \bar ss \rangle$ are gluon and u/d-quark and s-quark condensates, respectively. The functions~$I[n,m]$, $I_1[\mathcal{F}]$,~$I_2[\mathcal{F}]$~and~$I_3[\mathcal{F}]$
are
defined as:
\begin{align}
 I[n,m]&= \int_{m_c^2}^{\mathrm{s_0}} ds \int_{m_c^2}^s dl~ e^{-s/\mathrm{M^2}}\,l^n~(s-l)^m ,\nonumber\\
I_1[\mathcal{F}]&=\int D_{\alpha_i} \int_0^1 dv~ \mathcal{F}(\alpha_{\bar q},\alpha_q,\alpha_g)
 \delta^{\prime}(\alpha_{\bar q}+ v \alpha_g-u_0),\nonumber\\
 I_2[\mathcal{F}]&=\int D_{\alpha_i} \int_0^1 dv~ \mathcal{F}(\alpha_{\bar q},\alpha_q,\alpha_g)
 \delta(\alpha_{\bar q}+ v \alpha_g-u_0),\nonumber\\
 I_3[\mathcal{F}]&= \int_0^1 du \, \mathcal{F}(u), 
 \end{align}
 where $\mathcal{F}$ represents the corresponding photon DAs and the measure ${\cal D} \alpha_i$ is defined as

\begin{eqnarray*}
\label{nolabel05}
\int {\cal D} \alpha_i = \int_0^1 d \alpha_{\bar q} \int_0^1 d
\alpha_q \int_0^1 d \alpha_g \delta^{\prime}(1-\alpha_{\bar
q}-\alpha_q-\alpha_g).~\nonumber\\
\end{eqnarray*}
 
%\end{widetext},

%%%%%%%%%%%%%%%%%%%%%%%%%%%%%%%%%%%%%%%%%%%%%%%%%%%%%%%%%%%%%%%%%%%%%%%%%%%%%%%%%%%%%%%%%%%%%%%%%%%%%%%%%%%%%%%%%%%%%%%%%%%%
%%%%%%%%%%%%%%%%%%%%%%%%%%%%%%%%%%%%%%%%%%%%%%%%%%%%%%%%%%%%%%%%%%%%%%%%%%%%%%%%%%%%%%%%%%%%%%%%%%%%%%%%%%%%%%%%%%%%%%%%%%%%%%%%%%%%%

\end{widetext}

\bibliographystyle{elsarticle-num}
\bibliography{Singlycharm2.bib}

\end{document}